\documentclass[3p,times,procedia]{elsarticle}
\usepackage{nupha_ecrc}

\volume{00}

\firstpage{1}

\journalname{Nuclear Physics A}

\runauth{}

\jid{nupha}

\jnltitlelogo{Nuclear Physics A}

\usepackage{amssymb}
\RequirePackage{lineno}
\usepackage[figuresright]{rotating}

\begin{document}

\begin{frontmatter}



%% Instructions from Editor: Please use the following \dochead only in the preprint version (e-print arXiv etc.); 
%% use empty \dochead{} when submitting to Nuclear Physics A!
\dochead{XXVIIIth International Conference on Ultrarelativistic Nucleus-Nucleus Collisions\\ (Quark Matter 2019)}
%% e.g. \dochead{17th International Conference on Dynamical Processes in Excited States of Solids}

\title{Measurement of the charge separation along the magnetic field with Signed Balance Function   in 200 GeV Au + Au collisions at STAR}


\author{Yufu Lin  for the STAR Collaboration   \fnref{xxx} } %  %\ead{yufulin@mails.ccnu.edu.cn}
%\ead{yufulin@mails.ccnu.edu.cn}
\fntext[xxx]{yufulin@mails.ccnu.edu.cn}

\address{ Key Laboratory of Quark \& Lepton Physics (MOE) and Institute of Particle Physics, Central China Normal University, Wuhan 430079, China}

\address{Brookhaven National Laboratory, Upton, New York 11973, USA}

%\begin{linenumbers}
 
\begin{abstract}
 Experimental searches for Chiral Magnetic Effect (CME) in heavy-ion collisions have been going on for a decade, and so far there is no conclusive evidence for its existence. 
Recently, the Signed Balance Function (SBF), based on the idea of examining the momentum ordering of charged pairs along the in- and out-of-plane directions, has been proposed as a probe of CME.  
In this approach, a pair of observables is invoked: one is $r_{\mathrm{rest}}$, the out-of-plane to in-plane ratio of $\Delta B$ measured in pair's rest frame, where $\Delta B$ is the difference between signed balance functions;  The other is a double ratio, $R_{\mathrm{B}} = r_{\mathrm{rest}}/r_{\mathrm{lab}}$,  where $r_{\mathrm{lab}}$ is a measurement similar to $r_{\mathrm{rest}}$ but measured in the laboratory frame. 
These two observables give opposite responses to the CME-driven charge separation compared to the background correlations arising from resonance flow and global spin alignment.  Both  $r_{\mathrm{rest}}$ and $R_{\mathrm{B}}$ being larger than unity can be regarded as a case in favor of the existence of CME. It is found experimentally that $r_{\mathrm{rest}}$,  $r_{\mathrm{lab}}$ and $R_{\mathrm{B}}$ are larger than unity in Au+Au collisions at 200 GeV, and larger than realistic model calculations with no CME implemented.  These findings are difficult to be explained by a background-only scenario. 
\end{abstract}

%\end{linenumbers}

\end{frontmatter}

% \linenumbers

% \linenumbers
%\begin{linenumbers}  

\section{Introduction}
\label{intruduction}

It has been pointed out that the hot and dense matter created in relativistic heavy-ion collisions may form metastable domains where the parity and
time-reversal symmetries are locally violated, creating fluctuating, finite topological charges~\cite{ref1}. In non-central collisions, when such domains are immersed  in the ultra-strong magnetic fields produced by spectator protons,  they can induce electric charge separation parallel to the system's orbital angular momentum --- the chiral magnetic effect (CME)~\cite{ref2}. 
To study the CME experimentally one has to look for the enhanced fluctuation of charge separation in the direction perpendicular to the reaction plane, relative to the fluctuation in the direction of reaction plane itself. This is the basis of all CME searches in heavy-ion collisions.
Recently,  a new method,  namely the Signed Balance Function (SBF) method, is proposed as an alternative way to study the charge separation induced by CME in relativistic heavy-ion collisions~\cite{tang2019probe}.
The SBF method is based on the idea of examining the fluctuation of net momentum ordering of charged pairs along the in- and out-of-plane directions.  In this approach, a pair of observables were proposed,  one is $r_{\mathrm{rest}}$, the out-of-plane to in-plane ratio of $\Delta B$ measured in pair's rest frame, where $\Delta B$ is the difference between signed balance functions; the other is a double ratio $R_{\mathrm{B}} = r_{\mathrm{rest}}/r_{\mathrm{lab}}$,  where $r_{\mathrm{lab}}$ is a measurement similar to $r_{\mathrm{rest}}$ but performed in the laboratory frame.
These two observables have positive responses to signal, but opposite, limited responses to known backgrounds arising from resonance flow and global spin alignment. In this proceedings, we review tests made for the SBF with toy models, and give an update on tests made with realistic models. Latter ones include combinations of background and signal with various strengths. After that we will show SBF results from Au + Au collisions at 200~GeV measured by the STAR experiment at RHIC. 

\vspace{-0.38cm}

\section{Results and discussion}
\label{results}

\vspace{-0.08cm}

\subsection{ Review on toy model studies and update on realistic model studies}
\label{simulation}
 The major challenge in CME searches is that backgrounds, in particular those related to resonance elliptic flow and global spin alignment, can produce similar enhancement of fluctuations with the CME signal in the direction perpendicular to the reaction plane \cite{tang2019probe, ref4}. The effects of both signal and backgrounds have been implemented in toy model simulations~\cite{tang2019probe}, and for the configuration details of toy model please see Ref~\cite{toymodel}. With the toy model, the responses of SBF observables can be studied  using various signal and background combinations, in a controlled and systematic way.

\begin{figure}[htbp!]
\vspace{-0.4cm}
%\hspace{-0.75cm}
\centering 
\begin{minipage}[b]{0.33\textwidth} 
\centering 
\includegraphics[width=1\textwidth]{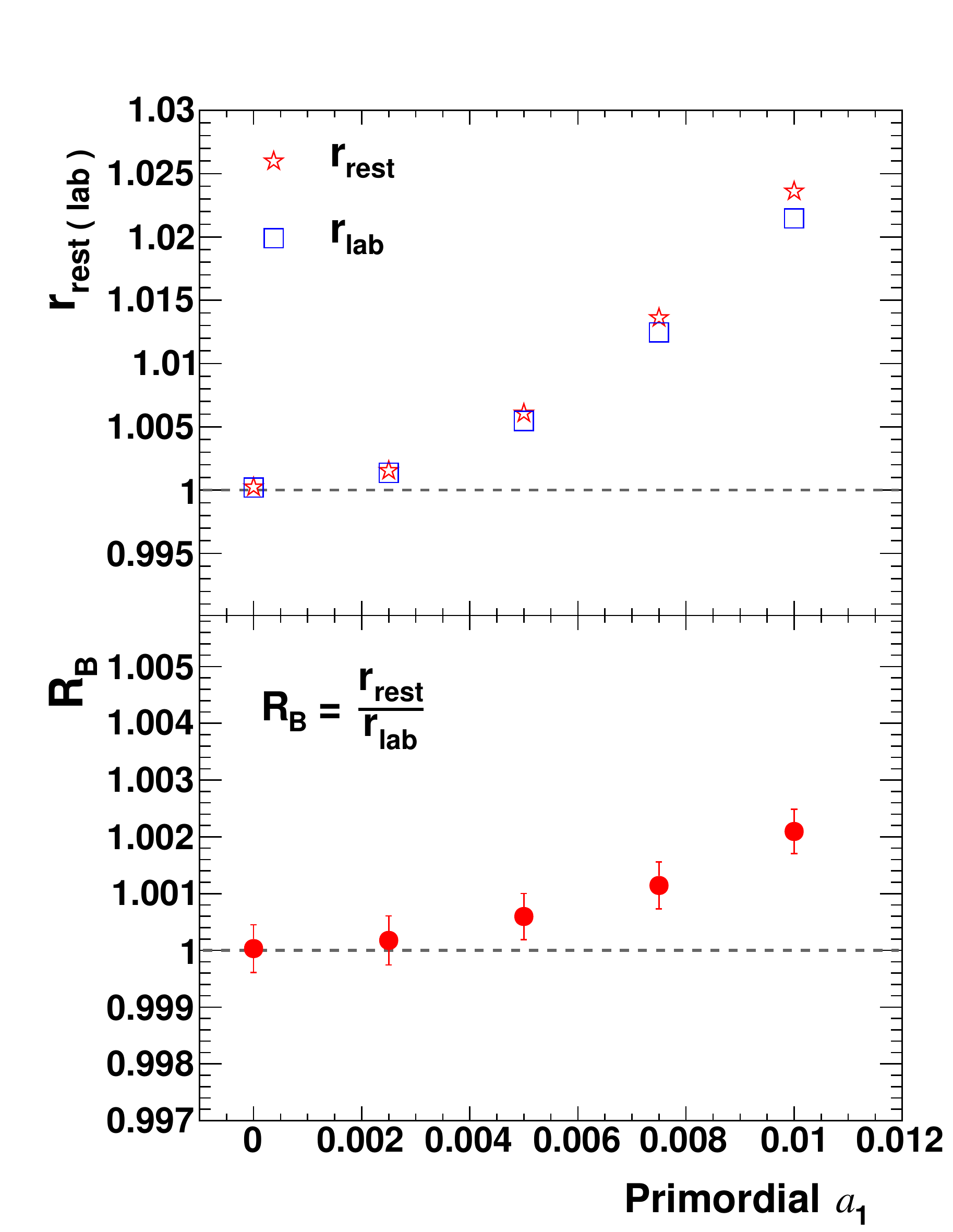}
\caption{ The $r_{\mathrm{rest}}$, $r_{\mathrm{lab}}$ and  $R_{\mathrm{B}}$  as a function of $a_{1}$ obtained for the toy model (signal only, no backgrounds) ~\cite{tang2019probe}. 
\vspace{3.4mm}
 }
\label{Fig.1}
\end{minipage}
\hspace{0.10cm}
\begin{minipage}[b]{0.33\textwidth} 
\centering 
\includegraphics[width=1\textwidth]{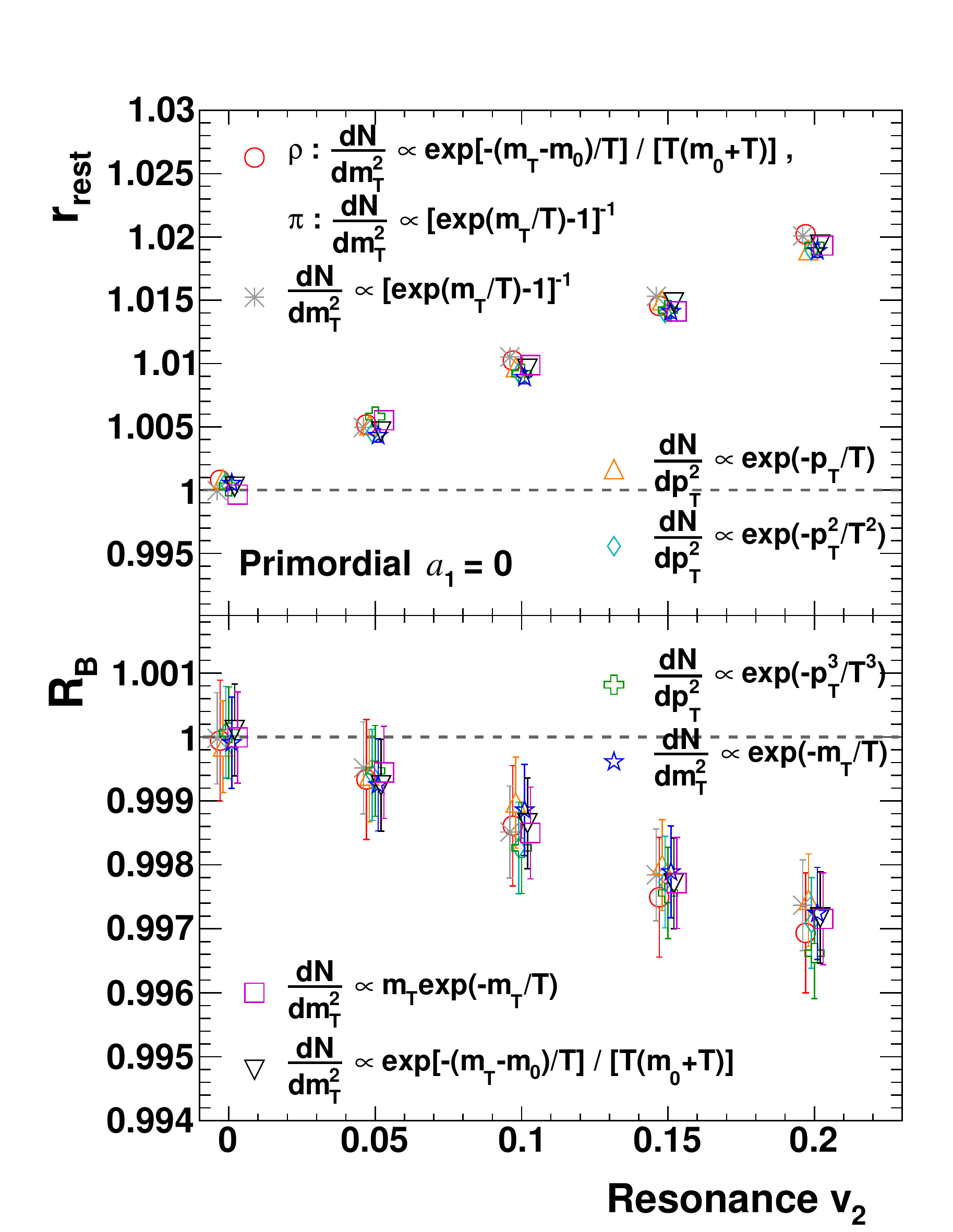}
\caption{  The $r_{\mathrm{rest}}$ and $R_{\mathrm{B}}$ as a function of resonance $v_{2}$ for various transverse momentum/mass spectra obtained for the toy model ~\cite{tang2019probe}. }
\label{Fig.2}
\end{minipage}
%\hspace{0.10cm}
%\begin{minipage}[b]{0.33\textwidth} 
%\centering 
%\includegraphics[width=1\textwidth]{fig_rho00Change_noFlowNoA1_compareSpectra.pdf}
%\caption{  $r_{\mathrm{rest}}$ and $R_{B}$ as a function of resonance $v_{2}$, for various $a_{1}$ values ( Fig comes from  ~\cite{tang2019probe}).}
%\label{Fig.3}
%\end{minipage}
\end{figure}

\begin{figure}[htbp!]
\vspace{-0.25cm}
%\hspace{-0.25cm}
\centering 
\begin{minipage}[b]{0.33\textwidth} %43
\centering 
\includegraphics[width=0.96\textwidth]{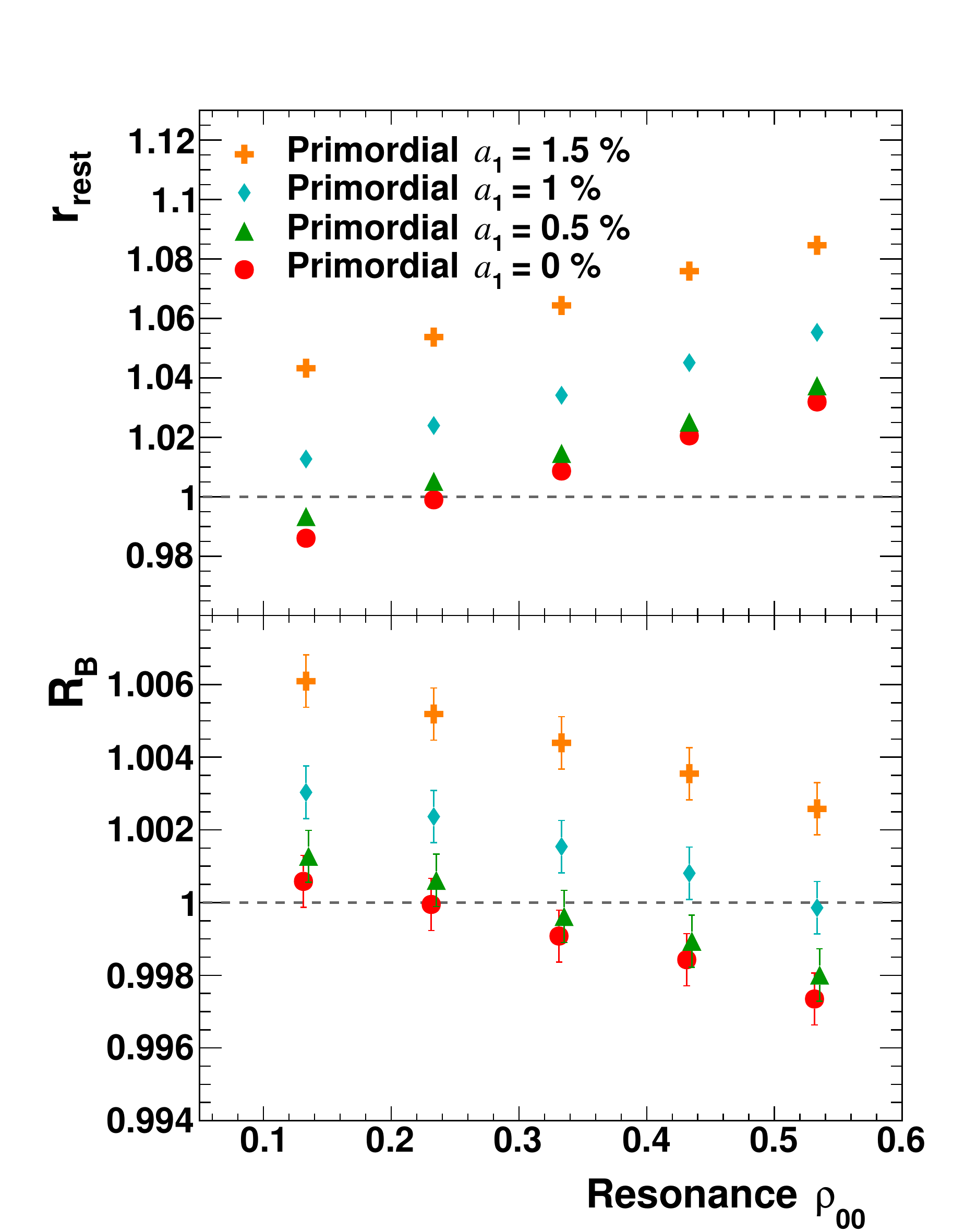}
\caption{ The $r_{\mathrm{rest}}$ and $R_{\mathrm{B}}$ as a function of resonance $\rho_{00}$ for various $a_{1}$ values obtained for the toymodel~\cite{tang2019probe}).}
\label{Fig.4}
\end{minipage}
\hspace{0.45cm}
\begin{minipage}[b]{0.33\textwidth} 
\centering 
\includegraphics[width=1\textwidth]{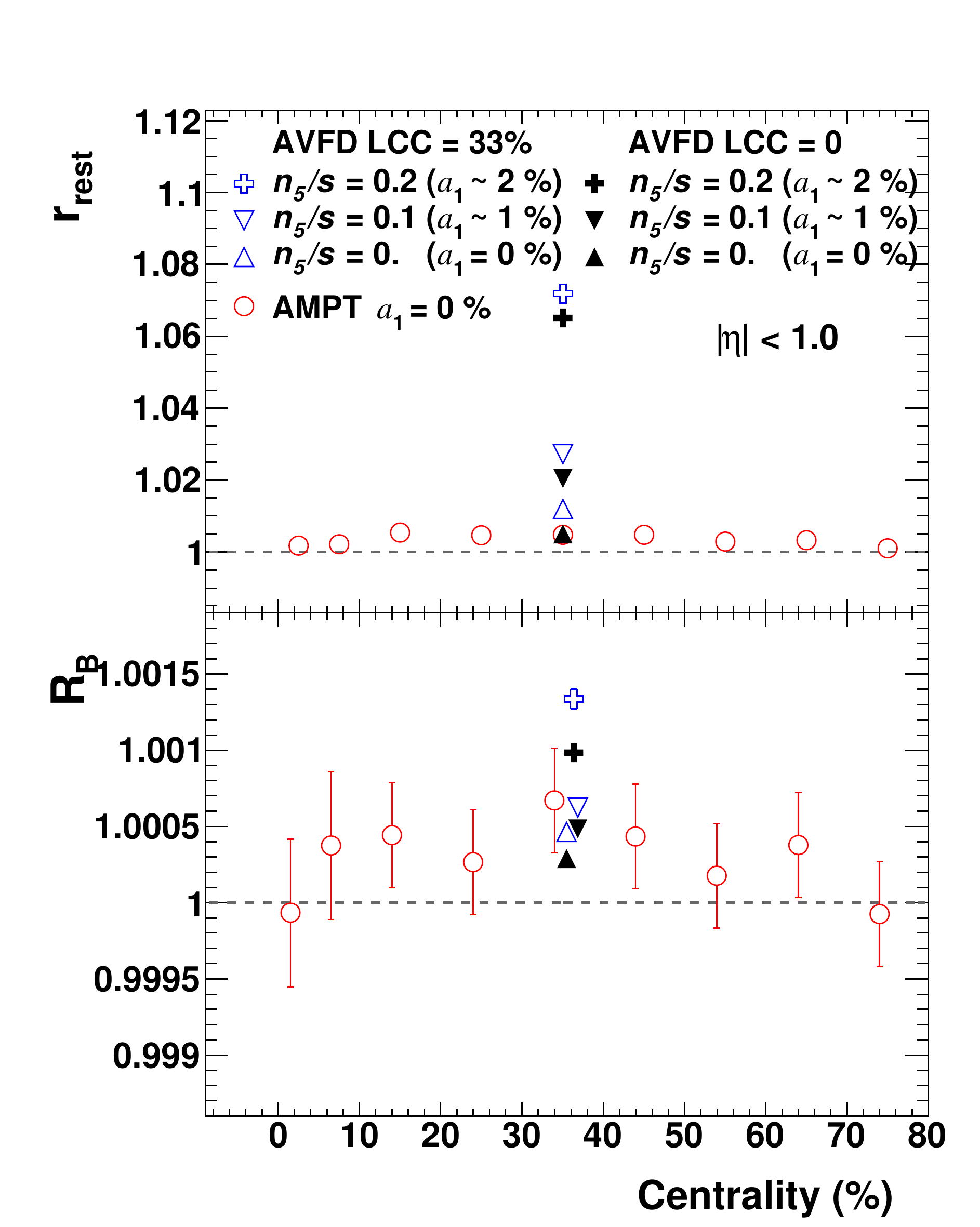}
\caption{The $r_{\mathrm{rest}}$ and $R_{\mathrm{B}}$ as a function of centrality, calculated for events from AMPT and AVFD. }
\label{fig:AMPT_AVFD}
\end{minipage}
\end{figure}

In Fig.~\ref{Fig.1} SBF observables are shown as a function of primordial $a_1$, which $a_1$ refers to the signal of CME, without any backgrounds. Here $a_1$ represents the strength of CME signal~\cite{tang2019probe}. The $r_{\mathrm{rest}}$, $r_{\mathrm{lab}}$ and $R_{B}$ are consistent with unity when $a_{1}=0$, and increase with increasing $a_{1}$. The $r_{\mathrm{rest}}$ and $r_{\mathrm{lab}}$ follow each other to the first order but $r_{\mathrm{rest}}$ responds to signal more than $r_{\mathrm{lab}}$ does, which is the information shown in the bottom panel.  The results indicate that SBF observables are sensitive to the CME signal. 
The influence of elliptic flow of resonances are shown in Fig.~\ref{Fig.2}. The $R_{\mathrm{B}}$  is found to decrease with the increasing of resonance $v_2$, while the $r_{\mathrm{rest}}$  increases with it. The two observables show opposite dependence on resonance $v_2$  assuming various transverse momentum ($p_{T}$) spectra shape.  More cases with additional background configurations can be found in Ref~\cite{tang2019probe}. 
%Similar simulation to study the effects of global spin alignment ($\rho_{00}$ can be found in ref~\cite{tang2019probe}, $r_{rest}$ and $R_B$ give a opposite responses to the CME and backgrounds. 
Figure ~\ref{Fig.4} shows toy model results with CME signal and two major backgrounds (resonance flow and global spin alignment) considered, which are  closer to the realistic scenario. One can see that similar to the case of resonance flow only, $r_{\mathrm{rest}}$  and $R_{\mathrm{B}}$ respond in opposite directions to the change of global spin alignment ($\rho_{00}$). On top of that, both increase with increasing signal ($a_1$). It will be a case supporting CME if both $r_{\mathrm{rest}}$ and $R_{\mathrm{B}}$ are larger than unity, barring additional background from Local Charge Conservation (LCC) and Transverse Momentum Conservation (TMC). Both LCC and TMC have to be studied with realistic models, which will be presented below.

\vspace{-0.18cm}

The two observables are also tested with two popular realistic models, namely  the AMPT~\cite{amptLin:2004}  and  AVFD~\cite{avfdref} models. Both models can reasonably describe data's key features (spectra,  elliptic flow, etc.).  For the AMPT version that is used in the test, no CME signal is implemented and charge-conservation has been assured. It can serve as a good baseline for apparent charge separation arising from pure backgrounds.
In the  AVFD model~\cite{avfdref}, the anomalous transport current from CME has been implemented by introducing finite ratio of axial charge over entropy ($n_{5}/s$),
%the anomalous transport current from CME has been implemented (the CME signal is sensitively related to the initial condition for the axial charge, $n_{5}/s$), 
allowing a  quantitatively and systematically study on observable's responses to signal embedded an environment of realistic backgrounds.  
Figure~\ref{fig:AMPT_AVFD} shows the results of  $r_{\mathrm{rest}}$ and $R_{\mathrm{B}}$ as a function of centrality for AMPT and AVFD events. To match the typical acceptance used by the STAR Collaboration, only particles in $|\eta|<1$ and $0.2 < p_{T} < 2$ GeV/$c$ are considered in the analysis.
For the two cases without  CME (AMPT and AVFD with $n_{5}/s = 0$), $r_{\mathrm{rest}}$ and $R_{\mathrm{B}}$ are consistent with unity within statistical uncertainties.  Both $r_{\mathrm{rest}}$ and $R_{\mathrm{B}}$ increase with increasing $n_{5}/s$ in AVFD results, results for the two LCC cases  (LCC =  33\% and LCC = 0\%).  The increase of the strength of LCC shifts both $r_{\mathrm{rest}}$ and $R_{\mathrm{B}}$ upwards, but only limited response is seen for the SBF observables when LCC changes from 0\% to 33\%. More detailed investigation on LCC is ongoing. 

\vspace{-0.08cm}

 \subsection{STAR's Results from Au + Au collisions at 200~GeV}
\vspace{-0.08cm}

\begin{figure}[htbp!]
	\centering
	\includegraphics[width=0.365\textwidth]{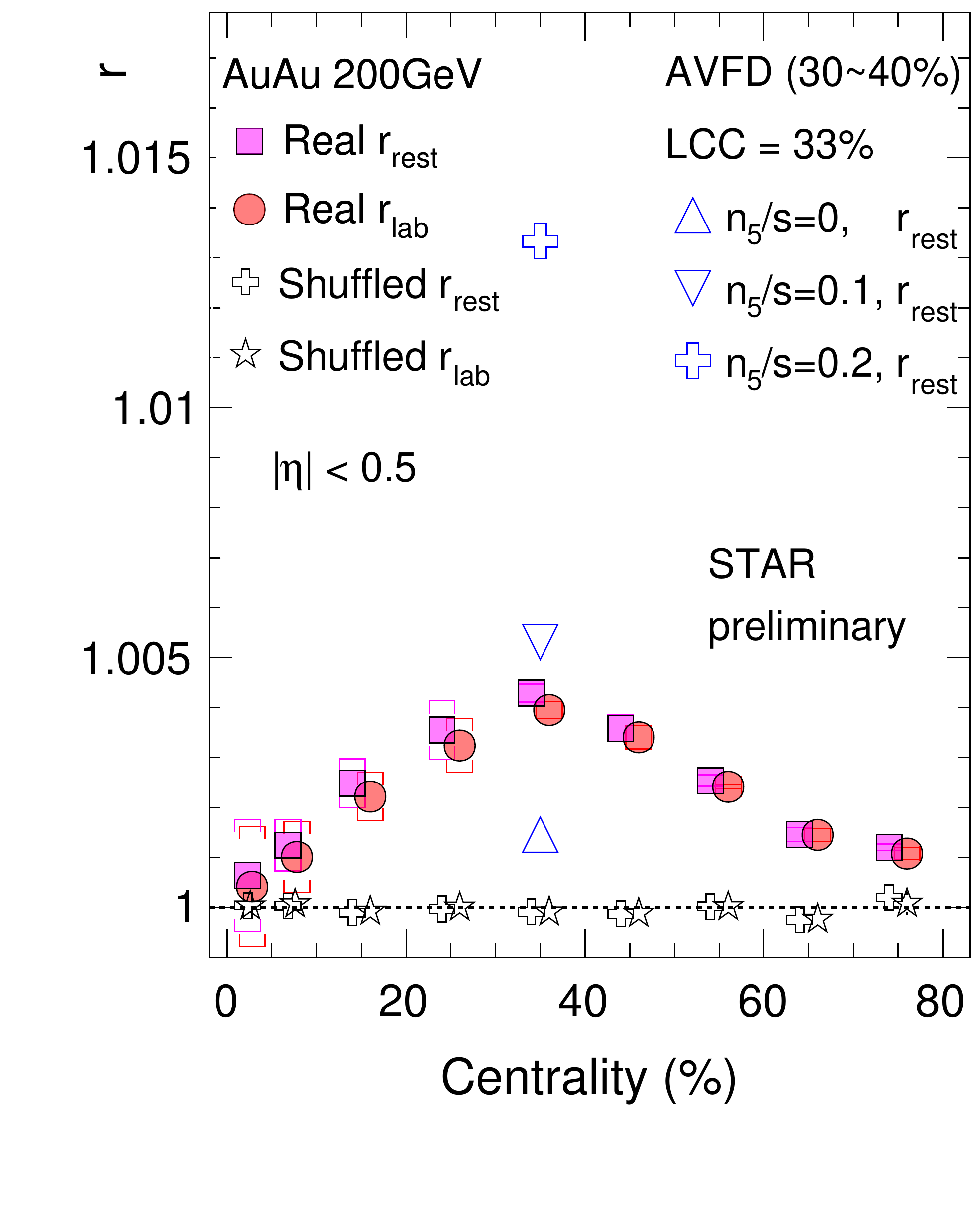}
	\includegraphics[width=0.4\textwidth]{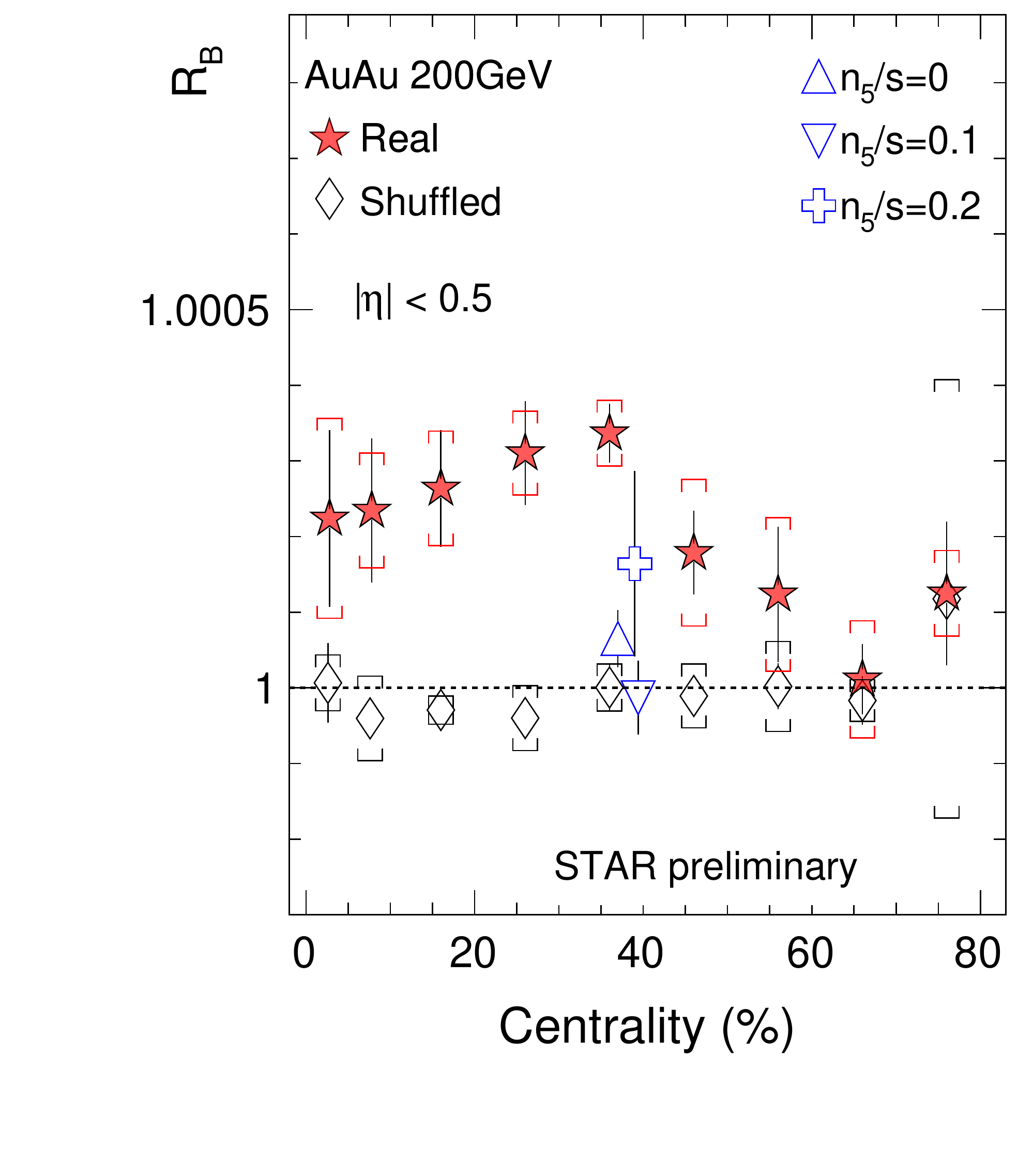}
	\vspace{-0.7cm}
	\caption{(Color online) 
	$r_{\mathrm{rest}}$ , $r_{\mathrm{lab}}$  and  $R_{\mathrm{B}}$ as a function of centrality from Au + Au 200~GeV at STAR.
	}
	\label{fig:finalresults}
\end{figure}

Experimental data used in this analysis are 200~GeV Au + Au collisions taken by the STAR  experiment in year 2016. About one billion minimum-bias events were used in the analysis.
The transverse momentum range for particles included in the analysis is $0.2 < p_T < 2 $ GeV/c. The second order event-plane  (EP), $\psi _{2}$, is reconstructed with Time Projection Chamber (TPC) withing $ 0.5< \eta <1.0$. Pions are used to calculate SBF, and they are identified with the information from both the TPC and the Time-Of-Flight detector. Pion kinematic region is confined in $|\eta| <0.5$, a different region than that for $\psi _{2}$ to avoid auto-correlation effects. In Fig.~\ref{fig:finalresults}, $r_{\mathrm{rest}}$ , $r_{\mathrm{lab}}$  and  $R_{\mathrm{B}}$ are shown as a function of centrality for both experimental data and AVFD model events. Results presented in Fig.~\ref{fig:finalresults}  are not corrected for the EP resolution. Instead, we smeared reaction plane in AVFD events with measured  event plane resolution in order to compare with data. The finite efficiency effect is also applied to AVFD events to assure a fair comparison.  One can find that both $r_{\mathrm{rest}}$, $r_{\mathrm{lab}}$, and $R_{\mathrm{B}}$ are larger than unity for all centralities for experimental data. As a consistency check, we also randomized each particle's charge while keep the total number of charged particles (positive and negative) in event unchanged. Such events and they are called shuffled events, and they are analyzed in the same way as what real events are analyzed. As shown in ~\ref{fig:finalresults}, SBF observables for shuffled events are at unity as expected. In the centrality of  30-40\%,  $r_{\mathrm{rest}}$ and $R_{\mathrm{B}}$ from data are both larger than the AFVD calculation without CME (the case of $a_1 =0$), indicating that there is a room to accommodate the CME explanation. Our overall observation is difficult to be explained by background-only model.

\vspace{-0.37cm}
\section{Summary}
\label{summary}
\vspace{-0.08cm}

We reviewed tests of SBF with toy models, and gave an update on studies made with two realistic models. 
Toy model simulation studies show that the two observables, $r_{\mathrm{rest}}$ and $R_{\mathrm{B}}$, respond in opposite directions to signal and backgrounds arising from resonance $v_2$ and $\rho _{00}$.  If both  $r_{\mathrm{rest}}$ and $R_{\mathrm{B}}$ are larger than unity, then it can be regarded as a case in favor of the existence of CME. 
In Au+Au collisions at 200~GeV,  $r_{\mathrm{rest}}$ ,  $r_{\mathrm{lab}}$ and $R_{\mathrm{B}}$ are found to be larger than unity, and larger than AVFD model calculation with no CME implemented. Our results are difficult to be explained by a background-only scenario. 

%The isobar analysis is still on going, we will not draw a conclusion until we finished all the system( including $UU$, pA).



%% References
%\vspace{-0.08mm}

%\section*{Acknowledgments}
%%%%%%%%%%%%%%%%%%%%%%%%%%%%%%%%%%%%%%%%%%%%%%%%%%%%%%%%%%%%%%%%%%%%%%%%%%%%%%%

\vspace{4mm}

\noindent$\bold{Acknowledgments}$
We thank  S. Shi and J. Liao for providing AVFD Beta1.0 source code. In particular we thank  the RHIC Operations Group and RCF at BNL. 
 Y. Lin is supported by the China Scholarship Council (CSC). 
This work is supported by the Fundamental Research Funds for the Central Universities under Grant No. CCNU19ZN019 and the Ministry of Science and Technology (MoST) under Grant No. 2016YFE0104800.

%\end{linenumbers}

\bibliographystyle{elsarticle-num}
%\bibliography{Reference}


\end{document}